# In-situ microwave characterization of ferromagnetic microwires-filled polymer composites: a review

F. X. Qin,[1] Y. Luo,[2] J. Tang,[1] H. X. Peng,[2] and C. Brosseau[3]

[1] 1D Nanomaterials Group, National Institute for Material Science,
1-2-1 Sengen, Tsukuba, Ibaraki 305-0047, Japan
[2] Advanced Composite Centre for Innovation and Science, Department of Aerospace Engineering,
University of Bristol, University Walk, Bristol, BS8 1TR, UK
[3] Université de Brest, Lab-STICC, CS 93837, 6 Avenue Le Gorgeu, 29238 Brest Cedex 3, France

This review describes the emerging research area and relevant physics of polymer-based composites enabled by amorphous ferromagnetic microwires. Fruitful results ranging from their tunable magnetic field and mechanical stress properties and influences of direct current on their microwave behavior are displayed in addition to the brief analysis on the underlying physics. The multifunctionalities exhibited strongly imply a variety of potential applications such as structural health monitoring and high-performance sensors. This article underlines that the future challenge mainly lies in proper microwire tailoring in expectation of a better microwave performance of microwire composites.

1. **Introduction**

Recent technological and industrial advances concerning magnetic sensors and devices require and anticipate materials with exceptional soft magnetic properties.[1,2,3] As a strong candidate, amorphous ferromagnetic microwires have been identified and intensively investigated for decades due to their excellent sensing properties. Their electromagnetic response can be conveniently tailored through designing the geometrical factors such as wire diameter and chemical composition. Moreover, they have been proved to have instant tunability in the millimeter wave frequencies towards external stimuli, e.g., external magnetic fields and mechanical stresses.[4,5] These merits, also evident in the giant magnetoimpedance (GMI) and giant stress-impedance (GSI) effects, provide a number of potential application areas such as microwave absorption, structural health monitoring, etc.[6] Ferromagnetic microwires are often categorized as Fe-based and Co-based, these elements giving good ferromagnetic properties when interacting with incident microwaves. To further enhance their magnetic properties, many efforts have been made in the microwire fabrication stage in addition to the subsequent physical and chemical wire treatments in the context of the needs of high-performance sensing devices.[4] However, this usually implied high costs, rendering such materials less suitable for industrial applications. Moreover, the Achilles' heel of free standing microwires is their small dimensions (usually tens of microns) and the fragile nature of their mechanical fracture behavior, which limits their applications. Despite the overall tendency of magnetic device miniaturization, it remains an issue that how we tackle the above problems without compromising their excellent sensing properties.

Most recently, a smart strategy of embedding microwires into polymer-based composites has been devised by our group to multifunctionalize the resultant composites, which enables additional electromagnetic functionalities and preserves the mechanical performance of the polymer matrix at the same time.[5] This proposition of microwire composites perfectly resolves the contradiction between the good magnetic properties and the difficulty of application deriving from the nature of microwires. Based on this perspective, fruitful experimental results and their underlying physics pertinent to ferromagnetic microwire composites have been systematically studied associated with other influential roles at play, for instance, the wire volume fraction,[7] wire length[8] and polymer matrices.[9] From all these exciting studies, this nascent research area is confirmed to have great scientific value and microwire composites are believed to be used in potential applications such as microwave devices and sensors in the future. It is therefore of significant interest to dispose a panorama on the microwire composites to summarize the on-going research on their fabrication, characterization and perspective application from the standpoint of engineering. The aim of this review is to present some of our recent progress on the fabrication and microwave characterization of microwire composites with or without external field/stress/current. With the understanding of the physics revealed, this versatile composite could be exploited to fit specific microwave applications and thus some perspectives can be envisaged.

The remainder of this review paper is organized as follows: we start by considering some fundamental aspects in Section 2 on the techniques for processing of microwires and their composites. Herein, different kinds of polymer composites were used as the matrices to offer mechanical constraint to microwires in order to study the particular external stimuli-dominated microwave behavior. Then in Section 3, we capitalize on recent advances on the microwave properties of microwire composites in presence of dc magnetic fields. A crossover field is identified in both Fe-based and Co-based wire composites. In addition, a double peak characteristic in the permittivity spectra of Co-based wire composites is discovered. Section 4 targets the stress effect of microwire composite, where the linearly increased permittivity dispersion is indicated in the Co-based composite while a contrary relation is obtained in its Fe-based counterpart. Section 5 is devoted to the influences of a direct current bias on the magnetoimpedance of polymer composites containing melt-extracted microwires. This current tunability can be exploited in specific engineered multifunctional microwire composites for active microwave devices. We conclude the whole review in Section 6 in addition to some remarks on the outlook.

## 2. Fabrication techniques of microwires and their composites and microwave characterization

Based on conventional techniques of fabricating amorphous alloys, a variety of fabrication techniques have been advanced for microwire processing, among which the modified Taylor-Ulitovskiy method is believed to be the most widely used approach.[10] Wires made by this technique usually have a metallic core with a glass coating closely attached. A detailed review discussing the fabrication and the GMI properties of microwires is listed here for interested readers.[4] In some of our studies, the microwires involved were processed by this technique.[11,12,13] The main advantages of such a technique are the repeatability of microwire properties in mass-production in a very economical way and the capability of fabrication continuous long pieces of microwire up to 10000 m.[3] In particular, by incorporating the glass-coated wires into polymer composites, interesting physics could be realized due to the introduction of the additional interfaces of glass-polymer matrix and wire-glass coating, taken into account that microwires are very sensitive to the external stresses on the surface.[14] On the other hand, alternative methods have been reported in the literature.[4] For instance, the melt-extraction (MET) technique has been recently adapted to yield high-quality and high-performance microwires with improved mechanical properties, as compared to other approaches such as in-water rotational spinning and glass-coated melt spinning.[15,16] The basic principle of MET is to apply a high-speed wheel with a sharp edge to contact the molten alloy surface and then to rapidly extract and cool a molten layer to be wires (Fig. 1).The MET technique has several advantages: (i) it has the highest solidification or cooling rate, which enables wires to be obtained with desirable amorphous structure; (ii) wires without a glass cover are more suitable for particular electronic packaging and sensor applications; and (iii) experimental parameters, i.e., linear velocity of the wheel, feed rate of the molten alloy, can be well controlled to ensure that the wires have uniform diameter and roundness.[17] However, the main drawback of this technique is the looseness of control on the wire diameter during the processing, rendering the typical wire geometry a semi-cylinder shape.[18]

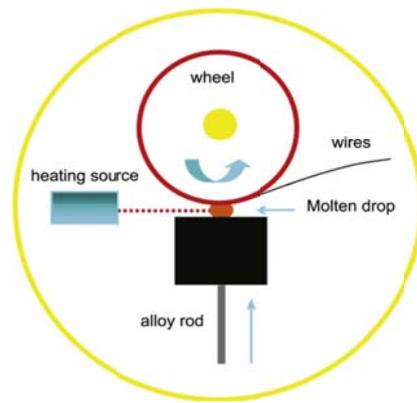

Fig. 1 Schematic view of melt-extraction wires processing. (reprinted with permission from Ref. 5, copyright 2013 Elsevier)

To aim at the microwire composites, different polymer materials should be considered to serve the particular investigation purposes as the candidate matrix material. Elastomers usually have much lower Young's modulus (2 MPa) compared to other polymers and hence would develop appropriate strain when subjected to even smaller external force. Therefore rubber composites containing microwires are suitable for characterizing their microwave response subjected to the external stresses. Regarding the experimental details, one piece of transparent silicone rubber sheets is used as the matrix material with parallelly arranged continuous microwires laid on, followed by bonding with another rubber piece using silicone glue to yield the resultant wire composites (Fig. 2).[19,20] Apart from the elastomers, epoxy is believed to be the most extensively studied material for all categories of composites and coatings for engineering applications due to its high mechanical properties. In this article, epoxy (Prime$^{TM}$ 20 or 27, purchased from Gurit, UK) is referred to as a non-conductive material and purely acts as the structural context to microwires, thus providing a good environment to investigate the overall field-tunable and current-tunable effects of wire composites.[8,7] The fabrication details may be described as follows: microwires were carefully mixed in a beaker with the epoxy resin and hardener (100:26 by weight) for 10 min to realize wire homogeneous dispersion. The well-stirred mixture, after degassing for 30 min in a vacuum oven at room temperature, was then cast into a rubber mould. It was subsequently cured at 50 °C for 16 h in an air oven. The samples for microwave characterization are parallelepipeds with length 70 mm, width 13 mm, and thickness 1.8 mm.

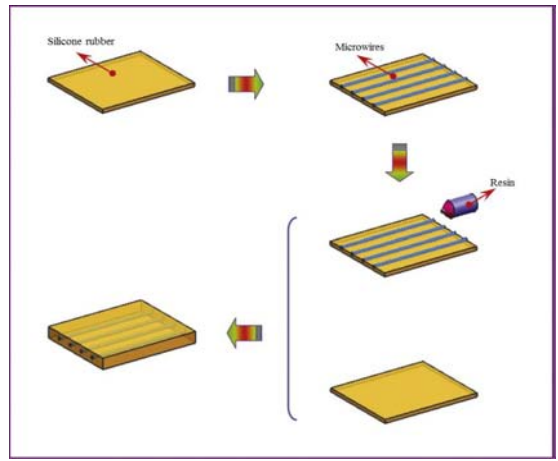

Fig. 2 Illustration of silicone rubber composites containing continuous microwires. (reprinted with permission from Ref. 5, copyright 2013 Elsevier)

The room temperature *S*-parameter measurement is carried out in a modified frequency-domain spectroscopy. The $S_{11}$ and $S_{21}$ parameters of an asymmetric microstrip transmission line containing the sample were measured by a vector network analyzer (Agilent, model H8753ES) with SOLT (short, open, load, and thru) calibration in the frequency band of 0.3 to 6 GHz under the quasi-TEM transverse electromagnetic mode.[21] The electromagnetic measurements were conducted with the wave vector perpendicular to the sample width direction. Using the Nicolson-Ross procedure for the transformation of the load impedance by a transmission line, $\varepsilon = \varepsilon' - j\varepsilon''$ is determined by the transmission $S_{21}$ and reflexion $S_{11}$ parameters.[21] An error analysis indicates modest uncertainties in $\varepsilon'$ (<5%) and $\varepsilon''$ (<1%) for the data. The measurement of $\varepsilon$ under external dc fields from 0 to 1000 Oe was performed by placing the microstrip line in between the poles of an electromagnet.[21] Our procedure for measuring $\varepsilon$ of soft materials under application of a uniaxial stress along the sample longitudinal direction is reminiscent of the innovative technique described in Ref. [22]. To study the current effect, a dc current is applied by connecting a current source to wires. To obtain accurate measurements of $\varepsilon$, it is particularly important to account for the residual air-gap between the sample and the line walls.[22]

### 3. Magnetic field tunable properties

It is established that Fe-based and Co-based wires have significantly different domain structures and hence rather different static and dynamic electromagnetic performances. As such, the field sensitivity of Co-based wires can be easily saturated at much lower magnetic fields. By looking into the field dependence of epoxy composites containing $Fe_{4.84}Co_{56.51}B_{14.16}Si_{11.41}Cr_{13.08}$ glass coated microwires, a larger magnetic field tunability (typically 150%) of the effective permittivity for the samples with microwire content of 0.026 vol. % is observed, together with an independent relation between field tunability and wire volume fraction.[7] This shifts our later investigations to the effect of wire length. Remarkably, similarities in the field dependence seen in the transmission and reflection coefficients for all samples investigated indicate that there is a crossover field at 300 Oe, i.e., the resonance frequency redshifts with fields increasing to 300 Oe then gradually blueshifts when fields are larger than 300 Oe (Fig. 3). We have elucidated that this crossover field is due to the competition between GMI effect and ferromagnetic resonance (FMR) of the microwires. At lower fields, the GMI effect dominates the dipole resonance behavior resulting in the compromise of $\varepsilon$ and reflection coefficients with increasing field due to the improved surface impedance.[7] Accordingly, transmission is enhanced. This explains the redshift of the reflection/transmission peaks at lower fields (Figs. 3(a) to 3(c)). On the other hand, the FMR prevails at higher magnetic bias, therefore inducing enhanced eddy current and skin effect because of the drastically mitigated skin depth, eventually contributing to the resonance peak blueshift (Figs. 3(e) to 3(f)).

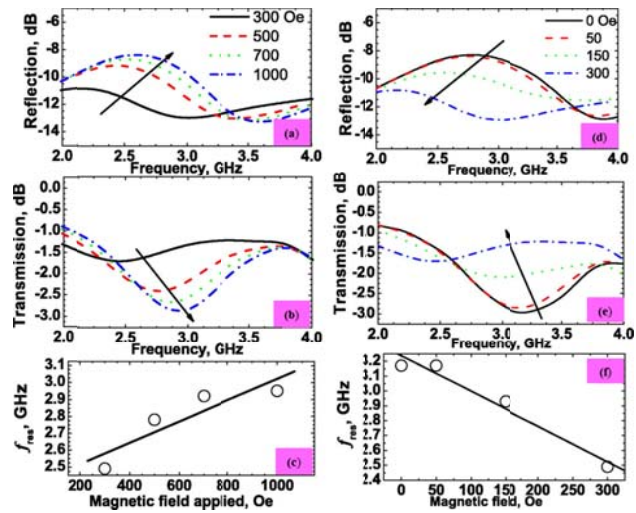

Fig. 3 The transmission spectra (a), reflection spectra (b), and field dependence of resonance frequency (c) of composites containing Co-based microwires under a dc field of 0-300 Oe. (d)-(f) Same as in (a) to (c) for the 300-1000 Oe. The solid line in (c) and (f) is a linear fit to the data. (reprinted with permission from Ref. 7, copyright 2012 AIP)

Notably, this crossover field is maintained in the composites containing Fe-based microwires but significantly increased to 600 Oe because of the much larger anisotropy field of Fe-based microwires (Fig. 4).[8] It is also indicated that for Fe-based wires the crossover field is dependent on the wire length, verified by the disappearance of such an effect when wire length is larger than 25 mm (not shown here), revealing the decisive role of wire length. This is because increasing the wire length to 35 mm would result in the wire entanglement after the oven curing process, making the dipole model no longer applicable, and thus jeopardize the permittivity tunability due to the dramatically reduced polarizability. From the microwave device point of view, it is of utmost importance to improve the field sensitivity when interacting with propagating electromagnetic waves. It is suggested that short wires in the composites are beneficial to obtain high field tunability for both Co-based and Fe-based microwires. One expects to obtain higher tunability by a further reduction of the wire length. However, during the curing process very short wires tend to aggregate and cause large magnetic loss due to additional anisotropy field induced by strong dynamic wire interactions.[23] Meanwhile, it has been reported that wire contacts would also decrease the threshold wire amount at which a percolation network is formed. Hence the wire interactions in the aggregates would also lead to undesirable high dielectric loss due to wave reflection rather than absorption.[24] Another possible approach would be to increase the wire content. However, one should note that a significantly higher wire concentration would also intensively increase permeability, and consequently reflection according to our earlier work.[7] These results provide strong impetus of designing glass-covered Fe-based and Co-based microwire-epoxy composites for adaptive materials for reconfigurable electronic devices and sensing applications where manipulation of overall dielectric properties can be managed via simply exerting a dc magnetic bias.

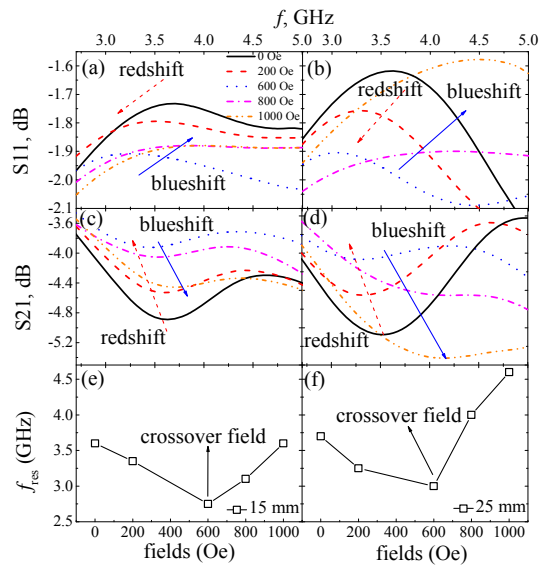

Fig. 4 Reflection coefficients, $S_{21}$, of Fe-based wire composites containing, respectively, (a) 15 mm and (b) 25 mm wires, transmission coefficients, $S_{21}$, of composites containing, respectively, (c) 15 mm and (d) 25 mm wires, and the field dependence of resonance position of composites containing (e) 15 mm and (f) 25 mm wires with external field up to 1000 Oe. (reprinted with permission from Ref. 8, copyright 2014 AIP)

Apart from the above glass coated wire containing composites, the field tunability in the context of different polymers containing MET wires was also investigated. For the rubber-based polymer matrix, samples containing randomly dispersed short-cut wires and continuous wires were studied.[25] Both of the samples express a strong dependence on the external fields. For the randomly distributed wire sample, the effective permittivity spectra of the composite sample filled with periodically arranged wires exhibit a double-peak feature in the measuring frequency range. The high frequency peak is associated with the GMI resonance and can be clearly seen at low magnetic bias whilst the resonance peak identified at low frequencies is found to be the dipole resonance peak.[25] Further investigation revealed why and how this two peak structure in the permittivity spectra is formed from the materials science point of view and gave a direct means of high resolution transmission microscope (HRTEM) to detect the intrinsic structure of microwires.[26] Assuming a core-shell (CS) structure of a microwire, it is considered that the high frequency resonance peak is likely to correspond to the nanocrystalline phase. It is confirmed by the HRTEM image that the nanocrystalline (Fig. 5(b)) and amorphous phase (Fig. 5(c)) separated by a clear transition region (Fig. 5(a)) identified along the radial direction from the wire surface to the inner core. The formulation of the nanocrystalline phase is attributed to the rapid cooling rate during the wire fabrication procedure. Furthermore, the electromagnetic simulation indicates that the Drude-Lorentz model can describe the effective permittivity dispersion quite well in the range of frequency explored.[26] Of particular note is that such a double peak effect does not exist in the thin glass-coated microwires tailored by the Taylor-Ulitovskiy method. This is due to the fact that such wires are generally too thin to realize a sufficiently high quenching rate which is of paramount importance for the formulation of nanocrystalline phase.[27] This work points to some interesting physics that the dielectric performance of composites containing MET wires can be magnetically tuned by external fields and hence suggests that this hybrid composite can be considered to be candidate material for highly sensitive magnetic sensors. The observation of a nanocrystalline phase on the wire surface fundamentally explains the double peak effect and is of practical importance for our future MET microwire composite design. From the above studies, the main advantages of microwire-polymer systems are twofold: (i) low filler volume fraction ensures miniaturization of future sensors; (ii) high sensitivity towards external dc fields. All these results offer essential guidelines for magnetic sensing application design.

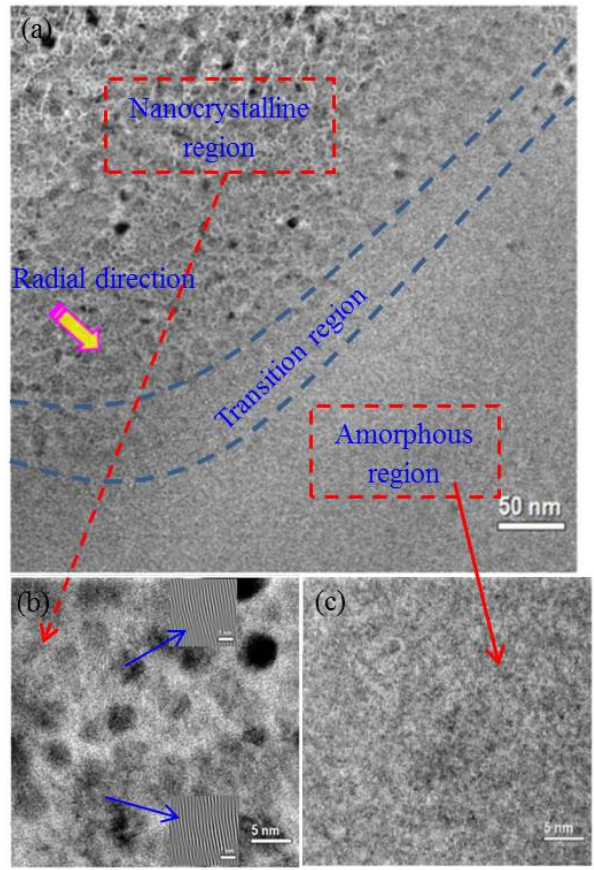

Fig. 5 (a) HRTEM image of microwires with a nanocrystalline, an amorphous, and a transition area. Local magnifications of (a) nanocrystalline region with their corresponding inverse transform selected area (IFFT) patterns and (c) the amorphous structure in (a). (reprinted with permission from Ref. 26, copyright 2013 AIP)

4. **Mechanical stress tunable properties**

Due to the stress effect on the impedance of amorphous wires, the stress will have significant impact on the prorogation of microwaves when they pass through the microwires. This is characterized by the variation of electromagnetic *S*-parameters with stress. This is the basic working principle for microwave NDT methods.[28] Very recently, the potential of using carbon fibers themselves as antenna and sensors to precisely detect the damage in carbon fiber reinforced polymers (CFRP) has also been demonstrated.[29] However, the interactions of CFRP and microwaves are limited by the high conduction loss of CF due to the high volume fraction. The use of microwires as sensor elements is advantageous for their strong interactions with microwaves, high sensitivity to external fields and thus low conductive loss. For composite containing ferromagnetic wires exhibiting GMI effect at microwave frequencies, the effective permittivity may depend on a dc magnetic field via the corresponding dependence of the surface impedance. The surface impedance can be changed by applying a stress which modifies the magnetic anisotropy and domain structure in wires. Thus, the effective permittivity may also depend on the external stress or strain. Following the stress tunable theory proposed by Panina,[30] we approached the strain effect on the electromagnetic responses from the technological aspects of the multifunctional composites.[8,19,20,23]

The Co-based glass-covered microwires were embedded in the silicone rubber in a parallel manner with spacing of 2, 1 and 0.8 mm respectively, followed by a standard curing process.[20] To investigate the stress effect, a mechanical load was applied along the sample longitudinal direction.[22] Other details of fabrication and characterization are stated in section 2. There is a significant influence of the microwire content on ε" (Fig. 5). For the sample with wire spacing of 2 mm, ε" is almost independent of the strain. For the 1 mm spacing sample, there is a remarkable dependence of ε" on strain, featured as an evolution of symmetric resonance peak at 4.5 GHz (Fig. 6(a)). As wire spacing decreases to 0.8 mm, the peak position and the evolution trend as a function of strain remains unchanged, although the shape becomes asymmetric (Fig. 6(b)). It is proved that strain dependence of the effective permittivity of microwire composites can be quantitatively analyzed using the Gaussian molecular network model (GMNM),[31] which implies a linearly increasing relationship between the stress tunability and the wire amount. The wire concentration dependence of ε is well described by the formula $\varepsilon = \varepsilon_m + 4\pi p \langle a \rangle$, where $p$ and $\langle a \rangle$ are respectively the wire volume fraction and the average polarizability.[32] The polarizability is primarily dependent on electric excitation. For the present composite configuration with microwires perpendicular to the electric field vector, although the sample is aligned with the tensile axis of the deformation apparatus, an axial component of electrical field still exists due to the inevitable misalignment of the wires within the sample and/or the possible inhomogeneity of the electrical field in the cell which may affect the electric excitation of the wires. This accounts for the observed dielectric response of the composite samples due to the polarization and the induced circumferential magnetization of the wires. For the magnetic microwires, the strain modifies the magnetization process and hence the circumferential permeability, resulting in the enhanced ε". All in all, it is indicated that with increasing wire concentration the composite presents increasing effective permittivity and strain sensitivity.

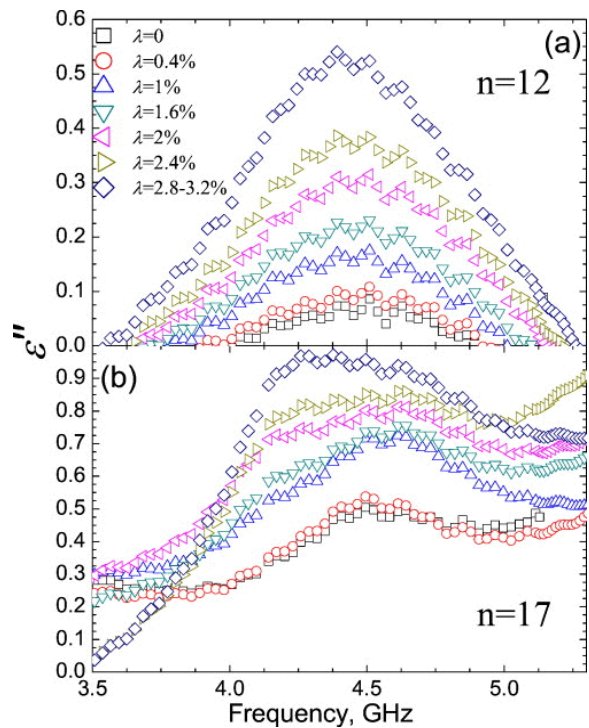

Fig. 6 Imaginary part of permittivity spectra of composites with varying amount of microwires (a) n=12 (wire spacing 1 mm) and (b) 17 (wire spacing 0.8 mm). (reprinted with permission from Ref. 20, copyright 2010 AIP)

From the application point of view, it is important to improve the stress sensitivity. Obviously, for composite samples containing Co-based microwires, this can be approached by increasing the microwire content of the composite. However, it should be noted that it does not necessarily mean the more the better. In this work, the sensitivity of permittivity to stress showed little change between samples of spacings of 1 and 0.8 mm. Surprisingly we find that the samples containing magnetic microwires and those containing nonmagnetic microwires appear to yield similar stress sensitivity of ε". This observation suggests an independence of the stress sensitivity in wire composites to the conductivity and magnetic permeability of the wires for a sufficiently high concentration of wires.

On the contrary, such positive correlation between stress tunability and wire amount does not hold for the composites containing Fe-based wires. To investigate the wire length effect, glass-covered $Fe_{77}Si_{10}B_{10}C_3$ microwires with different lengths of 15, 25 and 35 mm were sandwiched into the polymer matrix. For a sample with 25mm Fe-based wires, the magnitude of ε' and ε" exhibits an almost linear decrease as the strain is increased up to 2% (Fig. 7(a)) which contrasts with the stress-permittivity behavior for Co-based wire samples.[8] It was established that the dielectric response can be enhanced by improving circumferential permeability through modifying the surface plasmons.[33] Hence, the dielectric behavior basically arises from the dynamic coupling between the mechanical stresses and the circumferential anisotropy field. However, Fe-based wires have a negligible outer shell domain as opposed to Co-based wires.[4,34] Therefore, the coupling between the external stresses and the longitudinal anisotropy field of Fe-based wires is detrimental to the formulation of a circumferential field, thus decreasing the dielectric excitation to some extent. It appears that for 15 and 35 mm wires (Figs. 7(a) and 7(b)), the observed ε' is less intuitive. For the 15 mm wires, it is considered that the non-monotonic strain-tunability relation is due to the random orientation of microwires, i.e., it is more unfavorable to orientation of small wires. In this sense when further increasing external stresses we expect to recover the observed linear tunability vs. strain since the stress-anisotropy coupling increases significantly with large strain. Unfortunately, it was not possible to verify this conjecture because a partial breakdown appears for higher strain due to the fragile nature of polymer composites. For long wires, i.e., 35 mm, in addition to the mechanisms already discussed we should also consider the effect of the entanglement of the wires that could result in the significantly increased structure complexity in the microwire-composite system.

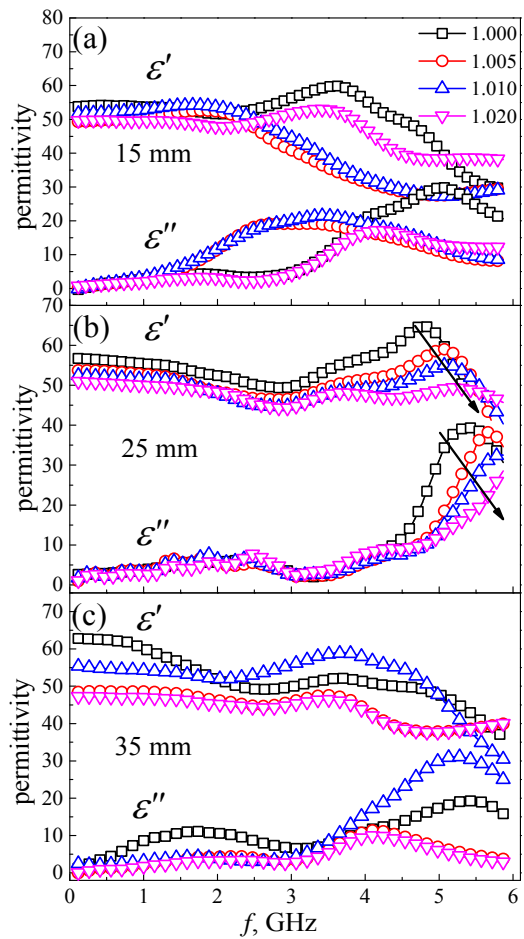

Fig. 7 The frequency dependence of the effective permittivity of composites containing Fe-based wires with length of: (a) 15 mm, (b) 25 mm, and (c) 35 mm, against strain up to 2.0%. The numbers in the inset indicate strain. (reprinted with permission from Ref. 8, copyright 2014 AIP)

On the other hand, an extended study demonstrates that a non-linear stress tunable effect is observed in the composite containing Co-based microwires over a broad range of strain due to the breakage of the wires in it (Fig. 8).[19] The analyses of permittivity and the strain sensitivity afford to identify two plausible mechanisms arising, namely, the stress and shape effects. When the strain exceeds 2.8%, breakdown of some wires occurs (inset of Fig. 8). Due to non-uniform interfacial conditions, each single wire experiences a different stress state resulting in partial but not total fracture of all wires. A recent study by Ipatov et al.[32] showed that the reduction of the wire length can significantly decrease the permittivity and leads to a blueshift of the resonance frequency. This explains why the present results differ from those reported above for linear tunability.[35] Thus, the observation here of broken wires is consistent with the decrease of the dielectric loss with stress of composites containing a smaller number of strained continuous wires. Some of the possible mechanisms which are involved to explain this observation are the stress effect, i.e., the stress changes the current distribution in the wires[36,30,37] and induces a higher dielectric loss and the shape effect, i.e., wires can be broken into shorter pieces and the anisotropy field of the wire becomes non-uniform, resulting in the reduction and the broadening of the resonance bandwidth. In addition, it is noted that the relevant relaxation mechanisms attributed to dipolar relaxation and Maxwell-Wagner-Sillars interfacial polarization[22,38] due to the boundaries separating microwires are expected to change due to wire fracture. Thus, the local properties of wires and the composite mesostructure are significantly altered. These studies have significant implications for research in the stress tunable microwire composites for sensing applications.

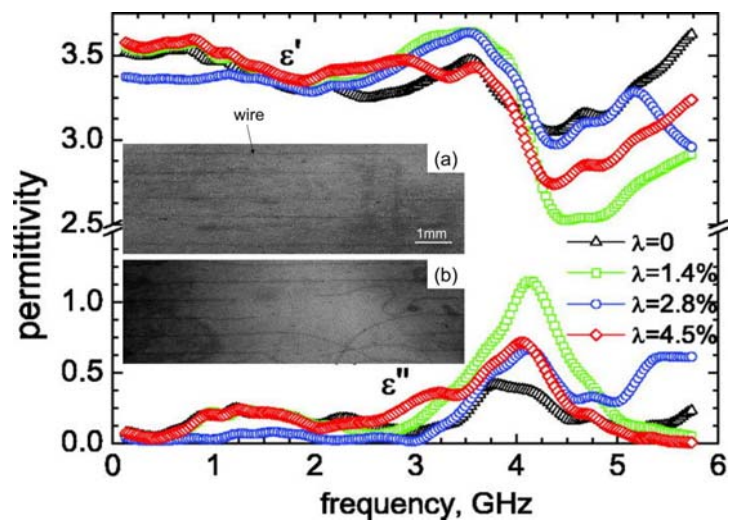

Fig. 8 The real, ε', and imaginary parts, ε", of the effective complex permittivity for different values of strain. The inset are images of the wire composite sample under uniaxial elongation corresponding to (a) λ=1.4%, i.e., elongated but not fractured wires, (b) λ=2.8%, i.e., showing broken wires. (reprinted with permission from Ref. 19, copyright 2011 AIP)

## 5. Current tunable properties

Applying a dc current offers an incisive opportunity to tune the microwave-wire interactions and, more importantly, the concept of microwire composite would make such material more competitive from application perspective. Most recently, we have reported the dc current (up to 100 mA) effect on the MI properties of composites containing Co-based MET wires.[17] A redshift of dielectric resonance frequency is noted together with a decrease in resonance peak of effective permittivity as the dc current increases with the presence of external magnetic bias of 500 Oe (Fig. 8). This resonance peak is identified as the FMR resonance peak of the microwires. Thus resonance peaks blueshifts to higher frequencies in the absence of dc current according to Kittel's equation.[39] On the other hand, the applied current would create an additional effective field which reflects the torque exerted on the axial magnetization by the field induced by the current perpendicular to the wire. This is, in turn, suggestive of a redshift of resonance peaks. From the permittivity amplitude peak versus dc current, it is noted that 10 mA has the minimum influence on the permittivity decrease. This is because the presence of a moderate current would realize stress relief resulting from Joule heating, which decreases the impedance and enhances the permittivity to a small extent. Increasing the dc current, the dominant effect of increasing the impedance prevails, thus decreasing the permittivity. However, it should be addressed that in further increasing the current amplitude to 50 mA the stress relief effect is expected to provide strongest compensation to the additional effective field and the permittivity becomes small again.

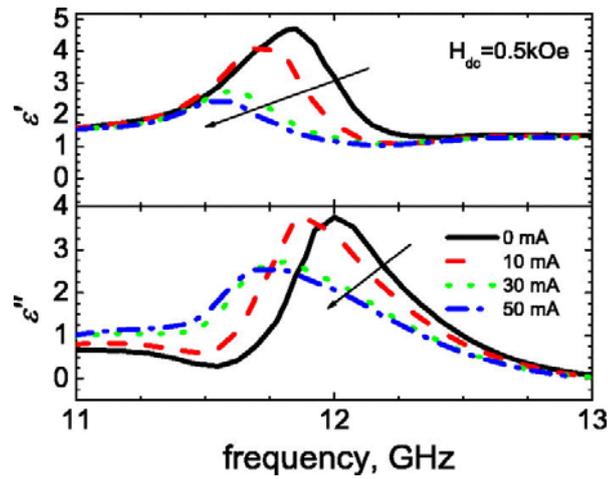

Fig. 8 (a) Real part of the complex effective permittivity spectra of the composite with a single microwire in the frequency range of 11–14 GHz with magnetic dc bias set to 0.5 kOe. The numbers in the inset indicate the bias current values. (b) As in (a) for the imaginary part. (reprinted with permission from Ref. 17, copyright 2014 AIP)

Finally, we point out that caution needs to be taken when the bias current is applied to avoid overheating of the microwire, or even burning it. Too large a current will also affect the microstructure of the metastable amorphous phase and possibly degrade its soft magnetic property. In that case, the bias current effect can be different from that discussed above since the magnetically hard surface will damage its tunability by external fields. This current tunability can be exploited in specific engineered multifunctional microwire composites for active microwave devices. Related work on the Fe-based microwire composites will be done very soon.

## 6.    Conclusions and outlook

This article reviewed fundamental and technological aspects of amorphous microwires enabled polymer composites and the primary conclusions are as follows: (i) by regulating the mesostructure and choosing different polymer composites as the matrix, the electromagnetic properties of the wire composites can be tailored and thus modulated to realize a variety of functionalities such as magnetic tunability, stress tenability, and current tunability; (ii) the microwave behavior of wire composites in presence of external fields indicates two remarkable features, i.e., the crossover field effect and double peak characteristic in the permittivity spectra; (iii) the strain-tunability of composites containing Co-based wires shows an increase trend of effective permittivity tunability with respect to moderate external stresses whilst an almost linear decrease trend is revealed for Fe-based wires enabled composites; (iv) dc current is shown to have profound effects on the electromagnetic manner of wire composites due to the two competing effects, that is, the current-induced additional effective field and stress relief arising from the Joule heating. These results indicate a versatile composite with the ability of all-around performance and capable of high sensitivity to external stimuli. The microwire composites are believed to be strong candidates for applications such as NDT, structural health monitoring, high sensitivity sensors, etc.

Using microwires as functional fillers has proved to be fruitful. A challenging question is how the electromagnetic properties of microwire composites can be further enhanced and tuned in a well-controlled manner. As mentioned above, the mesostructure of wire composites and the intrinsic electromagnetic properties of microwires are two decisive factors in determining the overall composite microwave performance. We have investigated the tunable properties of composites with wires randomly and periodically arranged. Naturally it is realized that their microwave response could be more controllable and sensitive to external stimuli providing moderate wire treatment is done before embedding into the matrix. Recently, it has been reported that several emerging techniques including the combined current-modulation annealing (CCMA),[40] twin zone Joule annealing[41] and multi-angle annealing[27] can significantly enhance the GMI effect and field sensitivity of Co-based microwires. Hence we believe that enhanced field/stress/current sensitivity in addition to other unique physics will be realized by performing such tailoring techniques to microwires prior to incorporating them into polymer context.

**Acknowledgements**

FXQ acknowledges JSPS fellowship and Grants-in-Aid for Scientific Research No. 25-03205. Yang Luo is supported from University of Bristol Postgraduate Scholarship and China Scholarship

Council. The authors would also like to thank Professor J. Gonzalez and A. Zhukov of Universidad del Pais Vasco, Spain for providing the glass-coated microwires.

**References:**

1. Arkadij Pavlovič Žukov and Valentina Žukova, *Magnetic properties and applications of ferromagnetic microwires with amorphous and nanocrystalline structure*. (Nova Science Publishers, 2009).
2. M Vázquez and AP Zhukov, Journal of magnetism and magnetic materials **160**, 223 (1996).
3. A Zhukov, J Gonzalez, M Vazquez, V Larin, and A Torcunov, Encyclopedia of nanoscience and nanotechnology **23**, 1 (2004).
4. M.H. Phan and H.X. Peng, Progress in Materials Science **53** (2), 323 (2008).
5. Faxiang Qin and Hua-Xin Peng, Progress in Materials Science **58** (2), 183 (2013).
6. Sergey N Starostenko and Konstantin N Rozanov, Progress In Electromagnetics Research **99**, 405 (2009).
7. FX Qin, HX Peng, J Fuller, and C Brosseau, Applied Physics Letters **101** (15), 152905 (2012).
8. Y Luo, HX Peng, FX Qin, BJP Adohi, and C Brosseau, Applied Physics Letters **104** (12), 121912 (2014).
9. Y Luo, HX Peng, FX Qin, M Ipatov, V Zhukova, A Zhukov, and J Gonzalez, Journal of Applied Physics **115** (17), 173909 (2014).
10. M Vazquez and A Hernando, Journal of Physics D: Applied Physics **29** (4), 939 (1996).
11. HX Peng, FX Qin, MH Phan, Jie Tang, LV Panina, M Ipatov, V Zhukova, A Zhukov, and J Gonzalez, Journal of Non-Crystalline Solids **355** (24), 1380 (2009).
12. F. X. Qin, H. X. Peng, N. Pankratov, M. H. Phan, L. V. Panina, M. Ipatov, V. Zhukova, A. Zhukov, and J. Gonzalez, Journal of Applied Physics **108** (4), 044510 (2010).
13. FX Qin, HX Peng, Z Chen, H Wang, JW Zhang, and G Hilton, Applied Physics A **113** (3), 537 (2013).
14. H. Wang, FX Qin, DW Xing, FY Cao, XD Wang, HX Peng, and JF Sun, Acta Materialia **60** (15), 5425 (2012).
15. Huan Wang, Dawei Xing, Xiaodong Wang, and Jianfei Sun, Metallurgical and Materials Transactions A **42** (4), 1103 (2011).
16. Huan Wang, Dawei Xing, Huaxin Peng, Faxiang Qin, Fuyang Cao, Guoqiang Wang, and Jianfei Sun, Scripta Materialia **66** (12), 1041 (2012).
17. FX Qin, J Tang, VV Popov, JS Liu, HX Peng, and C Brosseau, Applied Physics Letters **104** (1), 012901 (2014).
18. FX Qin, NS Bingham, H Wang, HX Peng, JF Sun, V Franco, SC Yu, H Srikanth, and MH Phan, Acta Materialia **61** (4), 1284 (2013).
19. Faxiang Qin, Christian Brosseau, and HX Peng, Applied Physics Letters **99** (25), 252902 (2011).
20. Faxiang Qin, HX Peng, C Prunier, and Christian Brosseau, Applied Physics Letters **97** (15), 153502 (2010).
21. Erwan Salahun, Patrick Queffelec, Le Floc'h, and Philippe Gelin, Magnetics, IEEE Transactions on **37** (4), 2743 (2001).
22. Christian Brosseau, Patrick Queffelec, and Philippe Talbot, J. Appl. Phys. **89** (8), 4532 (2001).
23. Yongjiang Di, Jianjun Jiang, Shaowei Bie, Lin Yuan, Hywel A Davies, and Huahui He, Journal of Magnetism and Magnetic Materials **320** (3), 534 (2008).
24. Zhihao Zhang, Chengduo Wang, Yanhong Zhang, and Jianxin Xie, Materials science & engineering. B, Solid-state materials for advanced technology **175** (3), 233 (2010).
25. FX Qin, C Brosseau, HX Peng, H Wang, and J Sun, Applied Physics Letters **100** (19), 192903 (2012).

26. F. X. Qin, Y. Quéré, C. Brosseau, H. Wang, J. S. Liu, J. F. Sun, and H. X. Peng, Applied Physics Letters **102** (12), 122903 (2013).
27. Jing‐Shun Liu, Da‐Yue Zhang, Fu‐Yang Cao, Da‐Wei Xing, Dong‐Ming Chen, Xiang Xue, and Jian‐Fei Sun, physica status solidi (a) **209** (5), 984 (2012).
28. K Arunachalam, VR Melapudi, L Udpa, and SS Udpa, NDT & E International **39** (7), 585 (2006).
29. Ryosuke Matsuzaki, Mark Melnykowycz, and Akira Todoroki, Compos. Sci. Technol. **69** (15-16), 2507 (2009).
30. DP Makhnovskiy, LV Panina, C Garcia, AP Zhukov, and J Gonzalez, Physical Review B **74** (6), 064205 (2006).
31. Christian Brosseau and Philippe Talbot, Meas. Sci. Technol. **16** (9), 1823 (2005).
32. Mihail Ipatov, V Zhukova, Larissa V Panina, and A Zhukov, PIERS Proc **5**, 586 (2009).
33. JB Pendry, AJ Holden, WJ Stewart, and I Youngs, Physical review letters **76** (25), 4773 (1996).
34. FX Qin, HX Peng, and MH Phan, Materials Science and Engineering: B **167** (2), 129 (2010).
35. FX Qin, N Pankratov, HX Peng, MH Phan, LV Panina, M Ipatov, V Zhukova, A Zhukov, and J Gonzalez, Journal of Applied Physics **107** (9), 09A314 (2010).
36. DP Makhnovskiy, LV Panina, and SI Sandacci, (Nova Science Publishers: Hauppauge, 2005).
37. LV Panina, SI Sandacci, and DP Makhnovskiy, Journal of Applied Physics **97** (1), 013701 (2005).
38. Christian Brosseau, Journal of applied physics **91** (5), 3197 (2002).
39. Charles Kittel, Phys Rev **73** (2), 155 (1948).
40. Jingshun Liu, Faxiang Qin, Dongming Chen, Hongxian Shen, Huan Wang, Dawei Xing, Manh-Huong Phan, and Jianfei Sun, Journal of Applied Physics **115** (17), 17A326 (2014).
41. JS Liu, FY Cao, DW Xing, LY Zhang, FX Qin, HX Peng, X Xue, and JF Sun, Journal of Alloys and Compounds **541**, 215 (2012).